\documentclass[]{JHEP3}

\usepackage{epsfig,multicol}

\newcommand{\Tr}{\mbox{\textrm{Tr}}}

\title{Contribution of disconnected diagrams to the hyperfine
splitting of charmonium}

\author{
The QCD-TARO Collaboration:
Philippe de Forcrand$^a$,   Margarita Garc\'{\i}a P\'erez$^b$,
Hideo Matsufuru$^c$,        Atsushi Nakamura$^d$,
Irina Pushkina$^e$,         Ion-Olimpiu Stamatescu$^f$,
Tetsuya Takaishi$^g$ and   Takashi Umeda$^h$ \\
$^a$Institut f\"ur Theoretische Physik,
    ETH-H\"onggerberg, CH-8093 Z\"urich, Switzerland,\\
    ~and Department of Physics, CERN, Theory Division, CH-1211 Geneva 23, Switzerland \\
    ~E-mail: \email{forcrand@phys.ethz.ch}\\
$^b$Instituto de F\'{\i}sica Te\'orica, Universidad Aut\'onoma de
     Madrid,\\
    ~Cantoblanco, 28049 Madrid, Spain\\
    ~E-mail: \email{Margarita.Garcia.Perez@cern.ch}\\
$^c$Computing Research Center, High Energy Accelerator Research
      Organization (KEK),\\
    ~Tsukuba 305-0801, Japan \\
    ~E-mail: \email{hideo.matsufuru@kek.jp}\\
$^d$IMC, Hiroshima University, Higashi-Hiroshima 739-8521, Japan\\
    ~E-mail: \email{nakamura@riise.hiroshima-u.ac.jp}\\
$^e$School of Biosphere Sciences, Hiroshima University,\\
    ~Higashi-Hiroshima 739-8521, Japan\\
    ~E-mail: \email{irina@riise.hiroshima-u.ac.jp}\\
$^f$Institut f\"ur Theoretische Physik, Universit\"at
        Heidelberg,  D-69120 Heidelberg, Germany,\\
    ~and FEST, Schmeilweg 5, D-69118 Heidelberg, Germany \\
    ~E-mail: \email{I.O.Stamatescu@ThPhys.Uni-Heidelberg.DE}\\
$^g$Hiroshima University of Economics, Hiroshima 731-0192, Japan\\
    ~E-mail: \email{takaishi@hiroshima-u.ac.jp}\\
$^h$Yukawa Institute for Theoretical Physics, Kyoto University,
    Kyoto 606-8502, Japan \\
    ~E-mail: \email{tumeda@yukawa.kyoto-u.ac.jp}
 }

\preprint{YITP-04-23 \\IFT-UAM-CSIC-04-14}

\abstract{
We discuss the contribution of  OZI-suppressed diagrams to the hyperfine
splitting of charmonium in lattice QCD. 
We study valence quark mass regions
from strange to charm quark masses.  No contribution of the
disconnected diagram is seen in the vector meson channel.
In the pseudo-scalar channel and for valence quark masses around 
the strange quark, the disconnected contribution induces a considerable 
increase of the meson mass. This contribution quickly decreases as the 
quark mass increases.  For charmonium the effect is very small
although a decrease of the pseudoscalar mass induced by the disconnected
contribution cannot be ruled out.}

\begin{document}

\section{Introduction}
 \label{sec:Introduction}

Lattice QCD has been able to provide a wealth of results for the hadron spectrum, 
both in the quenched approximation and with dynamical quarks \cite{reviews}.
Most of the low-lying hadron masses are reproduced under extrapolation to the
continuum limit and to the physical $u$ and $d$ quark masses.
Even in the quenched approximation there is a remarkable agreement with experimental 
data for the light hadron spectrum, deviations typically amounting to 10 
per cent.
The charmonium hyperfine splitting is, however, an exception to such success.
Recent systematic computations in the quenched approximation
have pointed out a significant discrepancy between the lattice results
and experimental data: the former is 30--40$\%$ smaller than the latter,
$\Delta M = M_{J/\psi}- M_{\eta_c} = 117$ MeV.

It is difficult to address the calculation of charmonium spectrum with
current computational resources. Since the charm quark mass is not well below the 
accessible lattice cutoffs, lattice formulations that reduce the $O(a m_c )$ 
lattice artifacts seem mandatory.
The most promising, while brute force approach, is to use
a relativistic formulation with sufficiently small lattice spacing
and $O(a)$-improved quark actions. This is the approach taken in
Ref. \cite{TARO03} by the QCD-TARO Collaboration, using  the 
nonperturbatively $O(a)$-improved Wilson quark action on quenched
isotropic lattices \cite{TARO03}.
Numerical simulations with lattice cutoffs
ranging from 2 to 5 GeV found $\Delta M =77(2)(6)$MeV
in the continuum limit. 

\FIGURE[t]{
\centerline{\epsfig{file=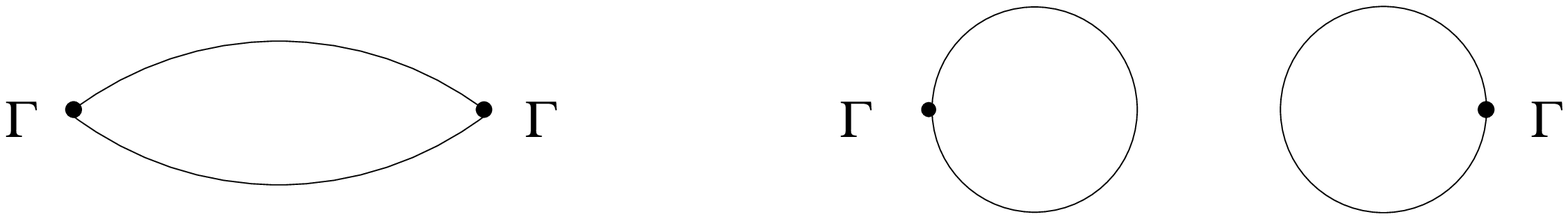,width=14cm}}
\vspace{-0.4cm}
\caption{
Connected (Left) and OZI-suppressed (Right) diagrams contributing to the
pseudoscalar ($\Gamma= \gamma_5$) and vector ($\Gamma= \gamma_\mu$) channels.
}
\label{fig:disc}
}

Other approaches involve the use of effective heavy quark actions. Among them,  
non-relativistic QCD (NRQCD) has been investigated most extensively.
The latest NRQCD quenched result is $\Delta M=55(5)$
MeV (with the scale set by the $P$-$S$ splitting) \cite{Trottier}.
In this case, lattice artifacts are difficult to control
since brute force elimination by taking the continuum limit is impossible.
Relativistic formulations, such as the Fermilab approach
\cite{EKM97,AKT03} and anisotropic lattices \cite{Klassen99,Chen01,Ume01,CPPACS02},
have an advantage in this sense. However, they give results for the,
continuum extrapolated, hyperfine splitting which also definitely 
deviate from experiment. Here it should be noted that for a quark mass 
not sufficiently smaller than the spatial lattice cutoff, the dynamics inside 
heavy quarkonia may not be precisely described by these actions whose spatial
derivative terms are defined only with nearest neighbouring sites.
This is because of the $O((a_\sigma p)^2)$ errors, where
$a_\sigma$ is the spatial lattice spacing and
$p$ the typical quark momentum. This error might become important for 
heavy quarkonium, for which $p\sim \alpha m_q$ \cite{CPPACS02,Aniso02a}.
This is in contrast with the situation for light and heavy-light hadrons,
where $p\sim \Lambda_{QCD}$. Therefore advantages of these relativistic 
formulations applied to heavy quarkonia are rather limited.

From all the previous studies, disagreement between the quenched
lattice calculation and experiment has been established
for the charmonium hyperfine splitting.

There are two candidates to explain this discrepancy.
The first one is dynamical quark effects.
Although a systematic study of these effects, involving a continuum extrapolation, 
has not been performed yet, several groups have
tried to estimate them \cite{SK01,El-Khadra00,DiPierro03}.
A recent lattice computation with 2+1 flavours of improved
staggered quarks at $a^{-1}\simeq 1.6$ GeV has reported 
a hyperfine splitting $\Delta M = 97(2)$ GeV \cite{DiPierro03}.
This value is still 20\% smaller than the experimental one.
Since they applied the Fermilab action with tadpole-improved
tree-level value for the clover coefficient, the remaining
discrepancy may be attributed to the $O(\alpha a)$ and
$O((ap)^2)$ systematic errors.
Systematic studies with higher lattice cutoffs and involving 
continuum limit extrapolations are strongly desired.

Another possible contribution, which has not been incorporated in any
of the lattice computations (quenched or unquenched) performed up to now, 
comes from OZI-suppressed (disconnected) diagrams as those in 
Fig.~\ref{fig:disc} \cite{TARO03}. Such diagrams must be included
in the evaluation of correlators of unflavoured mesons, such as $\eta_c$ 
and $J/\psi$.
Although, according to the perturbative picture, this contribution is expected 
to be small in heavy quarkonium, it might be non-negligible compared  
to the, also small, hyperfine splitting. Moreover, its effect might be 
enhanced. This happens, indeed,  in the light quark mass region, where
the contribution of the disconnected diagram to the pseudoscalar channel 
is strongly enhanced by the $U_A(1)$ anomaly \footnote{Note, however, that the
contribution of the $U_A(1)$ anomaly raises the mass of the pseudoscalar
while not affecting that of the vector. This effect would induce a decrease
of the hyperfine splitting, instead of the increase required to match the 
experimental value for charmonium.}. It is therefore important to quantify 
the size of such contribution to the charmonium correlator.

The goal of this paper is to examine the second possibility.
A similar analysis has recently been performed
by McNeile and Michael (UKQCD Collaboration) in Ref.~\cite{McNeile04}.
We aim here at an exploratory study to estimate
whether the disconnected diagram can give a significant
contribution to the charmonium hyperfine splitting.
This issue is not completely disentangled from the one 
of dynamical quark effects. Indeed, these effects can be particularly large for 
disconnected diagrams. In the quenched approximation closed quark loops, 
as those in the right panel of Fig.~\ref{fig:disc}, can only be connected 
through gluonic contributions.
For unquenched simulations, however, they can also be connected by insertion
of virtual quark loops.  
This induces a very different asymptotic behaviour between quenched and 
unquenched disconnected correlators, as discussed at length
for the case of the lattice determination of the $\eta'$ mass
\cite{Gusken}. In addition, through the disconnected diagram charmonium 
states mix with glueballs and lighter quarkonium. Mixing with the latter is 
completely neglected in the quenched calculation, while mixing with the 
former is only partially taken into account.  

We do not intend here to analyze
the failure of the quenched approximation in 
the calculation of the hyperfine splitting but to address the more general 
question of whether disconnected correlators 
can give a sizable contribution to this splitting.
Since it is expected that such contribution will be quickly obscured by
statistical noise, large statistics is essential for the present study
and this has dictated our choice of lattice parameters and lattice action.
The number of configurations collected for the quenched calculation of the 
hyperfine splitting in \cite{TARO03} is not sufficient for this study and 
increasing the statistics for these large lattices is beyond our computational 
resources. For this reason we use, instead, a large set (3200) of available 
configurations with a rather coarse $12^3\times 24$ lattice, with fixed lattice 
cutoff $a^{-1}\simeq 1.2$ GeV and with two dynamical flavours of staggered 
quarks with $a m_{\rm sea}=0.10$ (no attempt at a continuum extrapolation
will be presented in this paper).  
Given the rather large sea quark masses adopted, we expect
the effect of dynamical quarks to be relatively small, hence
our results give also an insight for the quenched situation. It would be 
important to study the sea quark mass dependence to
examine how such contribution shifts the charmonium masses for
physical sea quark masses and  $N_f=3$.

In order to treat charm quarks on our rather coarse lattice, 
we adopt the Fermilab quark action.
Since the disconnected diagram contribution to the
charmonium hyperfine splitting is hard to detect because of large
statistical noise, we also compute this contribution for lighter
valence quarks, and estimate how it varies as the valence quark mass
is increased towards that of the charm quark. This helps us determine
its value at the charm quark mass.
This is the only purpose of our varying the valence quark mass.
For lighter valence quarks, the inconsistency between our sea and valence
quark actions would generate systematic errors, which would cause
additional problems for the determination of the meson spectrum and of
the hyperfine splitting. We do not consider this regime here.

The paper is organised as follows.
In the next section we describe the set up for our calculation.
Section~\ref{sec:Numerical} presents our results. 
The last section is devoted to conclusions and discussion.

\section{Formulation}
 \label{sec:Formulation}

\subsection{Quarkonium correlation function}
 \label{subsec:correlator}

We define the quarkonia correlation function with a quarkonium 
operator $O(\vec{x},t)$, 
\begin{equation}
O(\vec{x},t)=\sum_{\vec{y}}\bar{q}(\vec{x}+\vec{y},t) 
\Gamma q(\vec{x},t) \varphi(\vec{y}),
\end{equation}
where $\varphi(\vec{y})$ is a smearing function and $\Gamma$ is
$4\times 4$ matrix which specifies the quantum numbers of the quarkonium state.

We take $\Gamma = \gamma_5$ and $\gamma_i$ for the
pseudoscalar and vector channels, respectively.
Since the operator $O(\vec{x},t)$ contains quark fields of 
the same flavour, by contracting the quark  lines
the correlator decomposes into a connected $C_{con}(t)$ and
a disconnected part $C_{dis}(t)$,
\begin{equation}
 C_{full}(t) = \sum_{\vec{x}}
 \langle O^\dagger(\vec{x},t) O(\vec{0},0) \rangle
 \equiv C_{con}(t) + C_{dis}(t).
 \label{eq:meson_corr}
\end{equation}
Using the quark propagator $D^{-1}(\vec{x},t;\vec{x'},t')$, where $D$ is
the Dirac operator, we can write  $C_{con}(t)$ and $C_{dis}(t)$ as
follows, 
\begin{eqnarray}
 C_{con}(t)&=&  -\sum_{\vec{x},\vec{y},\vec{z}}
              \left\langle \Tr [ \varphi(\vec{y})
{D^{-1}}^\dagger(\vec{x}+\vec{y},t;\vec{0},0)
\gamma_5 \Gamma \varphi(\vec{z}) D^{-1}(\vec{x},t;\vec{z},0) 
\Gamma \gamma_5] \right\rangle ,
 \label{eq:meson_corr_con}  \\
 C_{dis}(t)&=& \frac{1}{V_3} \left\langle 
      L(t)^\ast L(0)  \right\rangle,
 \label{eq:meson_corr_dis}
\end{eqnarray}
where $\Tr$ is the trace over the colour and spinor indices,
$V_3$ the spatial volume.
The quark loop diagram $L(t)$ is defined as
\begin{equation}
 L(t) =
  \sum_{\vec{x},\vec{y}} \Tr \left[
   \varphi(\vec{y})  D^{-1}(\vec{x},t;\vec{x}+\vec{y},t) \Gamma
 \right].
\label{eq:disc_diagram}
\end{equation}

As the valence quark mass increases, the disconnected correlator
is expected to decrease compared to the connected part
and the extraction of a signal for the disconnected part
may become increasingly difficult.
It is therefore crucial to improve the quarkonium operator
so as to increase the overlap with the ground state.
For this purpose, we use spatially extended operators
with smearing function $\varphi(\vec{x})$ in the Coulomb gauge.
For $\varphi(\vec{x})$ we adopt Gaussian
functions with width around the expected charmonium radius
and select the best one among them.

In order to evaluate the trace in Eq.~(\ref{eq:disc_diagram}),
we employ the complex $Z_2$ noise method \cite{Z2noise}.
This method is a popular technique to evaluate quark loop
contributions to correlators.
An application to the smeared operator, Eq.~(\ref{eq:disc_diagram}),
is straightforward.
We note that
one needs to solve the quark propagator only once for each noise vector 
for all the smearing functions applied.

\subsection{Valence quark action}
 \label{subsec:actions}

The coarse lattice we employ does not allow for an isotropic
formulation of heavy valence quarks whose mass is not sufficiently
smaller than the lattice cutoff. For this reason we adopt the Fermilab action
\cite{EKM97}, 
\begin{eqnarray}
 S_q &=& \sum_{x,y} \bar{q}(x) \left\{ \delta_{xy}
    - \kappa_\sigma \gamma_F \left[ (1\!-\!\gamma_4)U_4(x) 
      \delta_{x+\hat{4},y}      
      + (1\!+\!\gamma_4)U_4(x-\hat{4}) \delta_{x-\hat{4},y} \right] 
      \right.
      \nonumber \\
 & &  - \kappa_{\sigma} {\textstyle \sum_i}
     \left[ (r-\gamma_i)U_i(x) \delta_{x+\hat{i},y}
      + (r+\gamma_i)U_i(x\!-\!\hat{i}) \delta_{x-\hat{i},y} \right] 
       \nonumber \\
 & & \left.
    - \kappa_{\sigma} c_E {\textstyle \sum_{i}}
                                \sigma_{4i} F_{4i}(x,y) \delta_{x,y}
    - \kappa_{\sigma} c_B {\textstyle \sum_{i>j}}
                                \sigma_{ij} F_{ij}(x,y) \delta_{x,y}
  \right\} q(y), 
\label{eq:action_hopping}
\end{eqnarray}
where the spatial Wilson parameter is set to $r=1$.
The parameter $\gamma_F$ is to be tuned so that
the rest mass, $M_1\equiv E(\vec{p}=0)$, equals the kinetic mass,
\begin{equation}
  \frac{1}{M_2} \equiv \left. \frac{\partial^2 E(\vec{p})}{\partial p_i^2}
   \right|_{\vec{p}=0} ,
\end{equation}
for, for example, a meson dispersion relation.
We define $\kappa$ by incorporating tadpole improvement \cite{LM93}
as
\begin{equation}
 \frac{1}{\kappa} \equiv \frac{1}{u_0 \kappa_{\sigma}}
     - 2(\gamma_F + 3r - 4) \hspace{0.5cm} ( = 2(m_0+4) ),
\end{equation}
where $m_0$ is the bare quark mass \cite{Aniso01b}.
As the mean-field value of link variable, $u_0$, we adopt
the average value in the Landau gauge,
$u_0 = \langle \Tr U_\mu(x) \rangle /3$.
The value of $\gamma_F$ is tuned for each value of $\kappa$.
Then the chiral extrapolation is to be performed in $1/\kappa$,
a step that is not taken in this paper.
The clover coefficients, $c_E$ and $c_B$, control the 
$O(a)$ systematic uncertainty.
In this work, we adopt the tadpole improved values \cite{LM93},
\begin{equation}
  c_E = c_B = 1/u_0^3 .
\end{equation}

This action allows, in principle, for a relativistic treatment of heavy 
quarks but, as pointed out in Sec.~\ref{sec:Introduction},
even with nonperturbatively tuned 
$\gamma_F$, the results suffer from $O((ap)^2)$ errors,
where $p\sim \alpha m_q$ in the case of heavy quarkonium,
in addition to the $O(\alpha a)$ error from the clover terms.
We expect this effect not to be essential for a qualitative 
estimate of the size of the contribution from the
disconnected diagram. Better control over these errors 
is required, however, for a quantitative evaluation
and continuum extrapolation of the contribution.

\section{Numerical simulations}
 \label{sec:Numerical}

\subsection{Lattice setup}
 \label{subsec:Setup}

The numerical simulation is performed on lattices
of size $12^3 \times 24$, with two flavours of staggered
dynamical quarks.
The gauge action is the standard Wilson plaquette action
with $\beta=5.50$.
The dynamical staggered quark mass is $m_{sea}a=0.10$.
The configuration is generated with the hybrid R-algorithm
with $\delta t = 0.02$ and unit length of trajectory.
We prepare 32 independent initial configurations and generate
configurations in parallel.
Each configuration is separated by 5 trajectories, after 900
trajectories for thermalization.

The lattice cutoff scale is set by Sommer's hadronic radius,
$r_0$, which is defined through
$r_0^2 F(r_0)=1.65$, where $F(r)$ is the force between static quark
and antiquark \cite{Sommer94}.
Setting $r_0=0.5$ fm yields $a^{-1} = 1.2012(60)$ GeV.
For tadpole improvement \cite{LM93}, the mean-field value is
determined in the Landau gauge as $u_0 = \langle
\Tr U_\mu(x)\rangle/3$, giving 
$u_0 = 0.824285(89)$ and clover coefficient
$c_{SW} = 1.785$ for our lattice.
In the following analysis, the statistical error is estimated
by the jackknife method.

\subsection{Tuning of valence heavy quark action}
 \label{subsec:Tuning}

\FIGURE[t]{
\centerline{
\epsfig{file=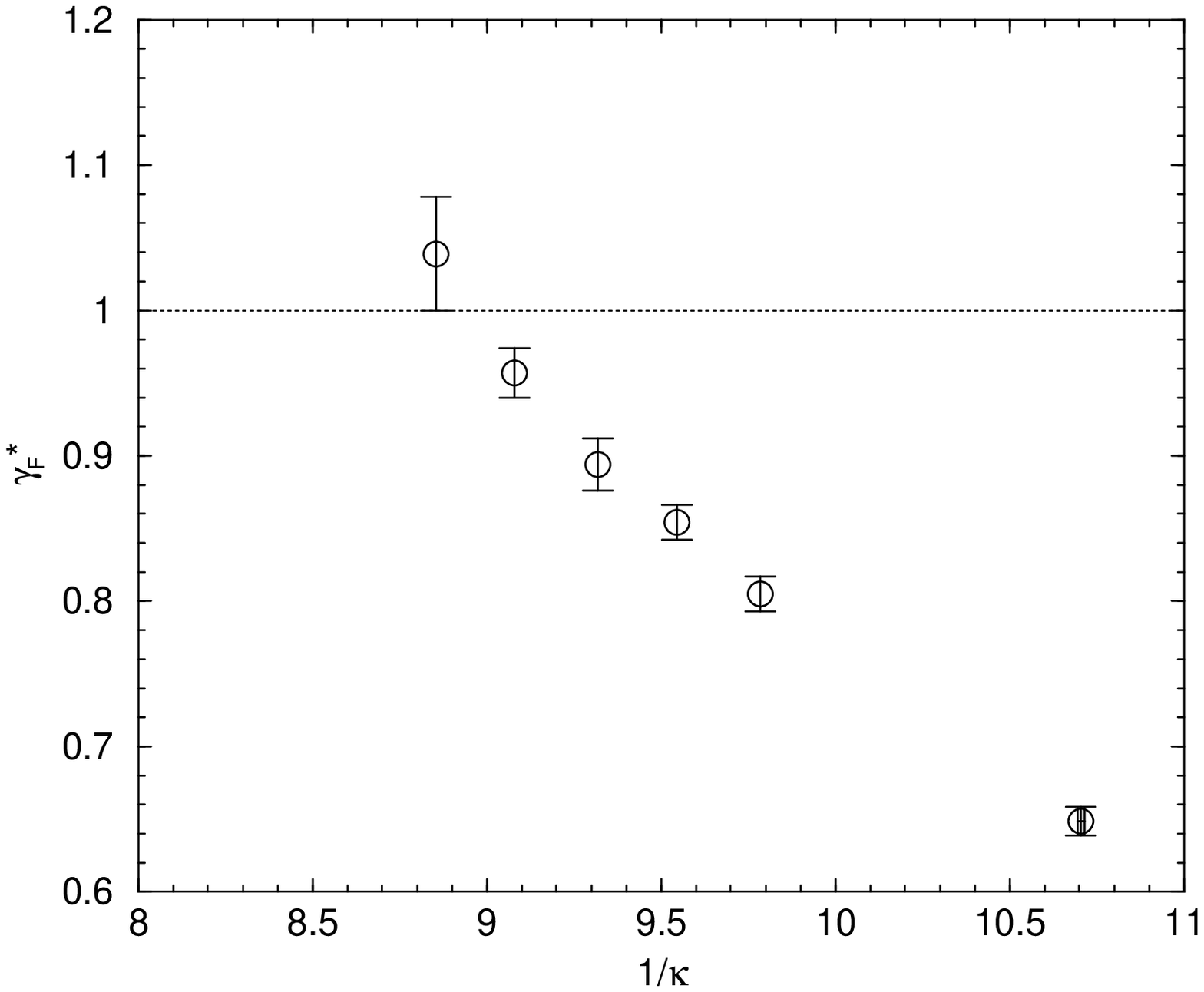,width=9cm}}
\vspace{-0.8cm}
\caption{
Tuned parameter $\gamma_F^*$ using the meson dispersion relation.}
\label{fig:calibration}
}

\TABLE[t]{
\caption{
Parameters of the valence quarks.
$N_{conf}$ is the number of gauge configurations used;
they are generated from 32 independent initial configurations.
The pseudoscalar and vector meson masses are
extracted from connected correlators with smeared operators.}
\label{tab:qparams}
\vspace{2mm}
\newcommand{\m}{\hphantom{$-$}}
\newcommand{\cc}[1]{\multicolumn{1}{c}{#1}}
\renewcommand{\tabcolsep}{1.3pc} 
\renewcommand{\arraystretch}{1.2} 
\begin{tabular}{cccccc}
\hline\hline
$\kappa$  &  $\gamma_F^*$ &  $N_{conf}$ & $N_{NV}$  &  $m_{PS} a$ & $m_V a$ \\
\hline
0.11294   & 1.039   &   1920    &    50     & 0.60714(92) &  0.8479(17)  \\
0.11013   & 0.957   &   1920    &   100     & 0.90071(81) &  1.0753(13)  \\
0.10732   & 0.894   &   1920    &   100     & 1.17518(74) &  1.3135(11)  \\
0.10476   & 0.854   &   1920    &   150     & 1.41022(70) &  1.5257(10)  \\
0.10220   & 0.805   &   3200    &   300     & 1.65156(51) &  1.75154(73) \\
0.093417  & 0.6485  &   3200    &   300     & 2.50132(48) &  2.56858(61) \\
\hline\hline
\end{tabular}
}

In this subsection, we describe the tuning of the parameter $\gamma_F$
in the valence quark action.
The tuning procedure is the same as for anisotropic lattices
\cite{Aniso01b}.
At each hopping parameter $\kappa$, the coefficient $\gamma_F$
in the quark action (\ref{eq:action_hopping}) is determined
nonperturbatively using the meson dispersion relation.
For the meson dispersion relation, we assume the relativistic form
\begin{equation}
  E(\vec{p})^2 = M_1^2 + \frac{M_1}{M_2}\vec{p}^{\ 2}
                     + O(\vec{p}^{\ 4}),
\end{equation}
where $M_1$ and $M_2$ are the rest and kinetic masses of
the meson.
The value of $\gamma_F$ is tuned so that $M_1=M_2$ holds.

The tuning of $\gamma_F$ is performed with connected correlators of
point operators,
namely $\varphi(\vec{x})=\delta(\vec{x})$, on 400 configurations.
The meson energies are fitted to a quadratic form for the
pseudoscalar and vector channels.
The obtained $\xi_F=\sqrt{M_2/M_1}$, the fermionic anisotropy,
are spin averaged and fitted to a linear function of
$\gamma_F$.
Interpolating to $(M_1/M_2)=1$, the tuned value of $\gamma_F$,
$\gamma_F^*$,
is determined.

The result for $\gamma_F^*$ is displayed in Fig.~\ref{fig:calibration}.
Since in the light quark mass region the Fermilab action smoothly
tends to the standard clover quark action, the value of $\gamma_F^*$
should approach unity as the quark mass decreases.   From
a tree level analysis,
it should be a decreasing function of $1/\kappa$, and in the light quark
mass region, a linear dependence in $m_q^2$ is expected
($m_q=(1/\kappa-1/\kappa_c)/2$,
where $\kappa_c$ is the critical hopping parameter (which is not
determined on the present lattice).
In Fig.~\ref{fig:calibration}, although the decreasing tendency
as $1/\kappa$ increases is indeed observed, the quark mass dependence
in the small mass region seems not in agreement  with the expected behaviour.
This is presumably due to large lattice spacing artifacts,
one of which apparently comes from the assumed form of the meson
dispersion relation.
This uncertainty may also cause $\gamma_F^*$ to approach a value
slightly different from unity \cite{Aniso01b}.
Although we will not try to correct these systematic errors 
for the present qualitative estimate of the hyperfine
splitting, it is essential to have them under control
in order to extract reliable quantitative determinations.


The value of $\kappa$ corresponding to the charm quark mass
is determined by interpolating the results for the vector meson
mass so as to reproduce the physical $J/\psi$ meson mass.
The value of $\gamma_F^*$ is also interpolated to that $\kappa$
value.
The resulting  $\kappa$ and $\gamma_F^*$ are listed in the last line
of Table~\ref{tab:qparams}, and displayed as the rightmost point
in Fig.~\ref{fig:calibration} together with the other five cases
listed in Table~\ref{tab:qparams}.

\subsection{Connected correlators}
 \label{subsec:Connected}

Charmonium correlators are computed for the quark parameters
listed in Table~\ref{tab:qparams}.
Let us start with the connected correlator, Eq.~(\ref{eq:meson_corr_con}).
In the following analysis, the error due to the uncertainty
in $\gamma_F^*$ is not evaluated.

We first observe the efficiency of the smearing technique on the
connected correlators.
Figure~\ref{fig:ep_con} displays the effective mass plot for
the connected correlators with local and smeared operators
in pseudoscalar and vector channels. Here the effective mass is defined
with 
\begin{equation}
\frac{C(t)}{C(t+1)}\equiv \frac{\cosh{[(T/2-t)m_{eff}(t)]}}
{\cosh{[(T/2-t-1)m_{eff}(t)]}}     .
\end{equation}
We use smearing functions with Gaussian form and several
width values, and select the one that gives better plateaus in 
the effective mass plot.
Figure~\ref{fig:ep_con} clearly shows that the smeared operator
considerably enhances the overlap with the ground state, which
dominates the connected correlator beyond $t=4$.

In Table~\ref{tab:qparams} we list the meson masses extracted from the
connected correlators. At the charm quark mass, the obtained hyperfine 
splitting, $\Delta M \simeq 81$ MeV, is consistent with previous results
\cite{DiPierro03}, and again far below the experimental value.

\FIGURE[t]{
\centerline{\epsfig{file=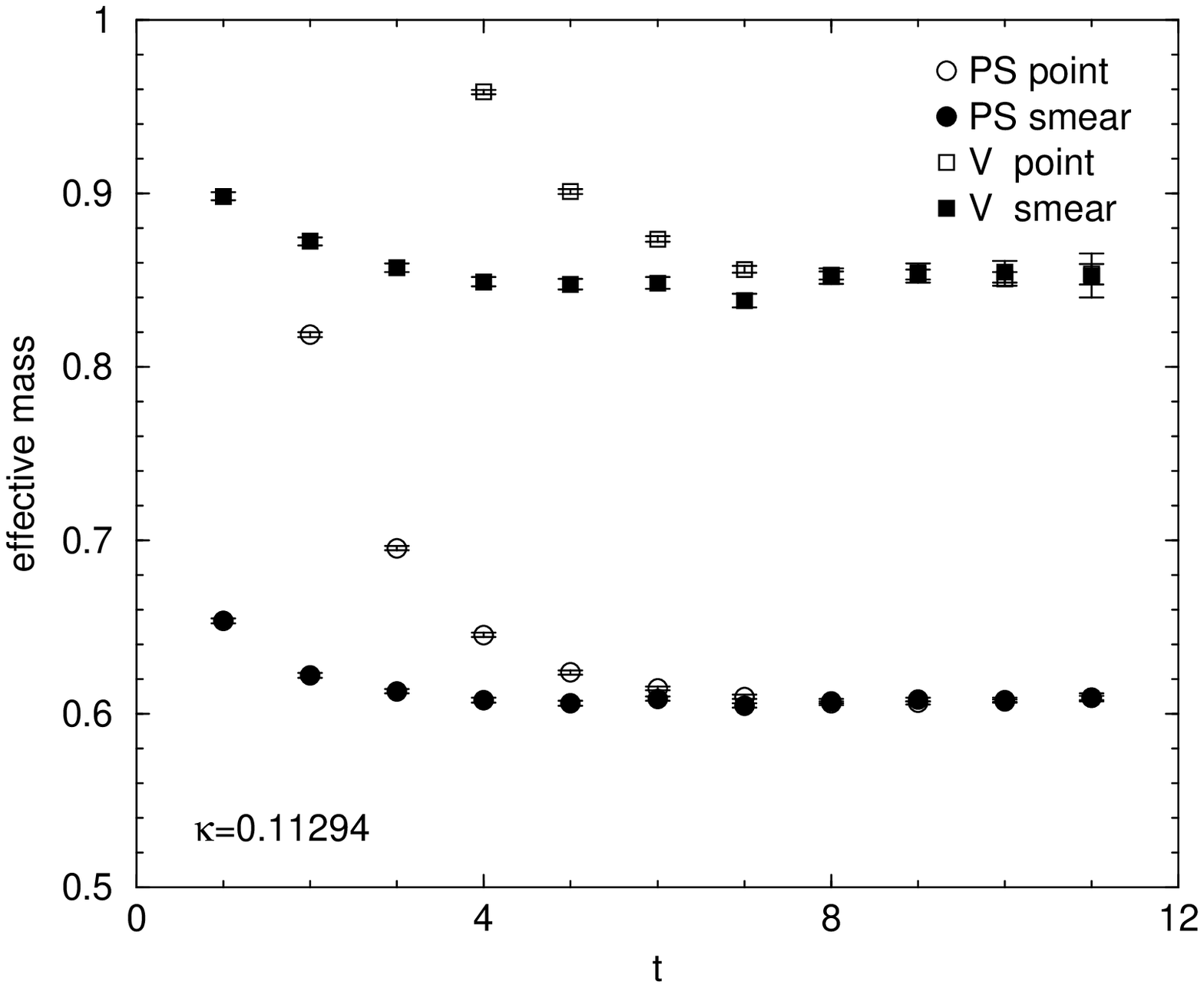,width=7cm} 
            \epsfig{file=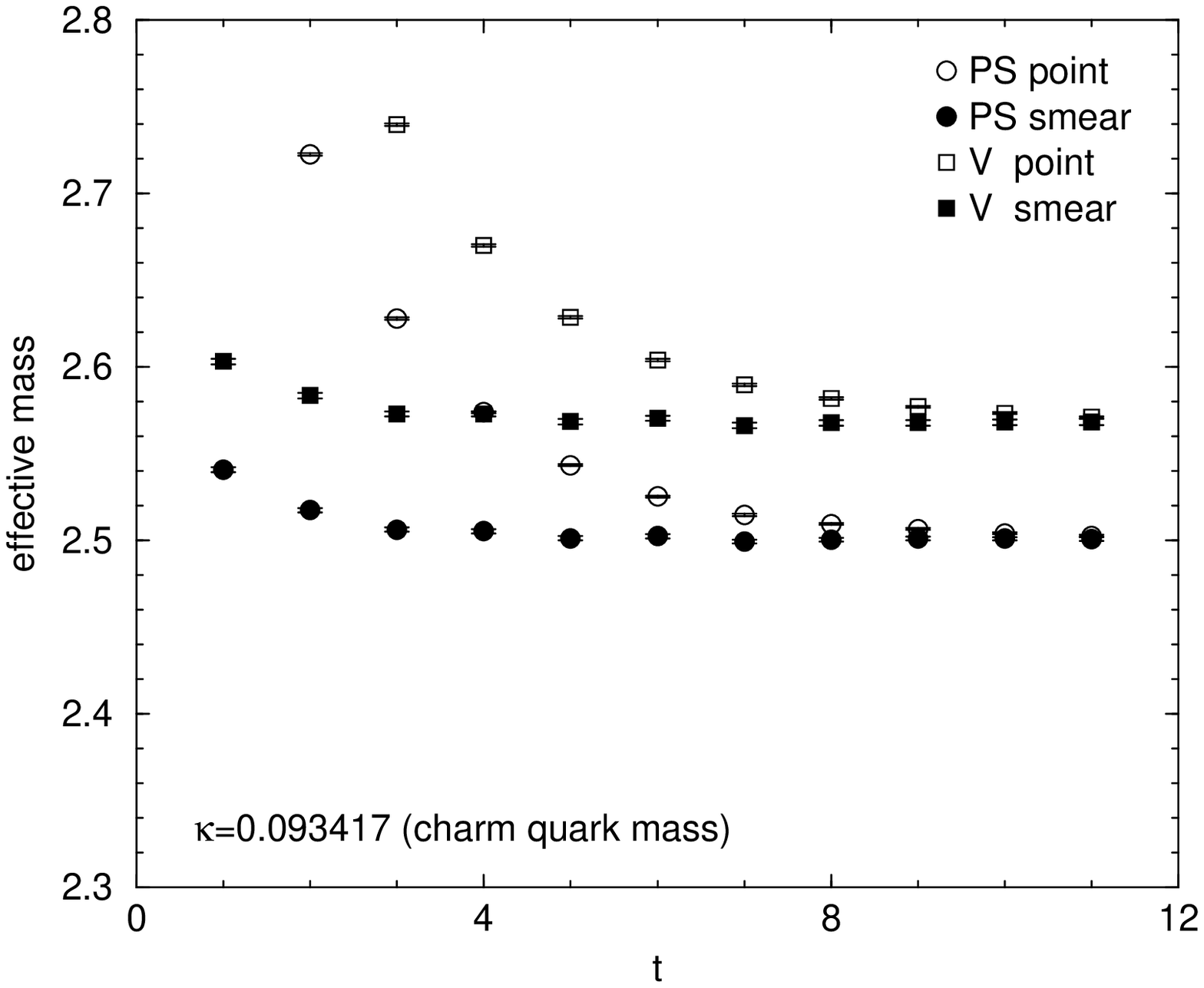,width=7cm} }
\vspace{-0.8cm}
\caption{
Effective masses of connected correlators with
local and smeared operators. }%
\label{fig:ep_con}
}

\subsection{Evaluation of disconnected diagram}
 \label{subsec:Disconnected}

\FIGURE[t]{
\centerline{\epsfig{file=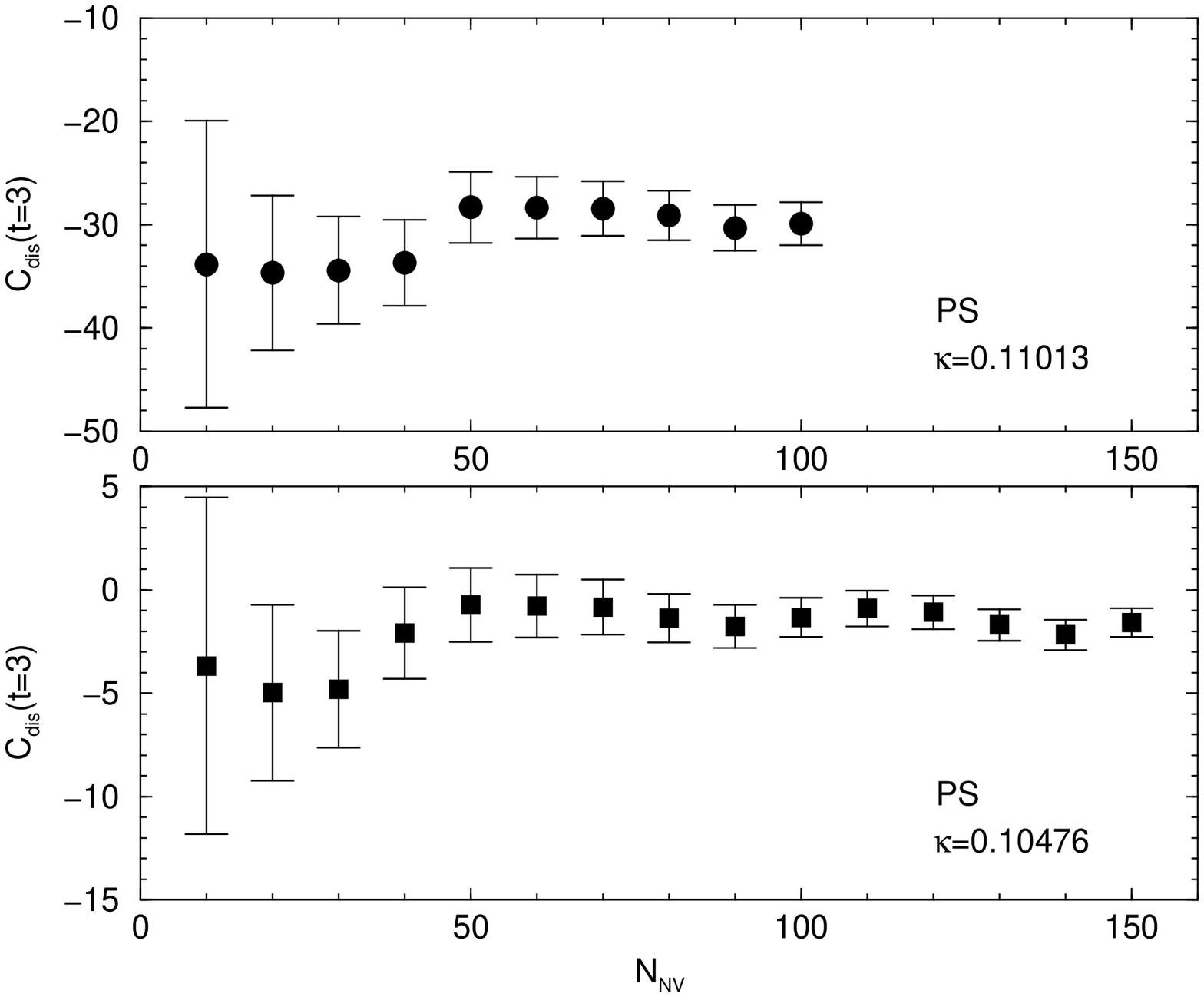,width=10cm} }
\vspace{-0.8cm}
\caption{
Typical samples of the dependence on the number of noise vector of
pseudoscalar disconnected correlators at $t=3$. Since the results are
averaged over configurations, the error includes both that of the  
noise method and that of the ensemble average.}%
\label{fig:Z2noise}
}

For an evaluation of the disconnected correlators,
Eq.~(\ref{eq:meson_corr_dis}), we apply the $Z_2$ noise method.
Figure~\ref{fig:Z2noise} shows typical samples of the pseudoscalar
disconnected correlator at a time slice $t=3$ as a function of
the number of noise vectors.
The top and bottom panels display the results for
$\kappa=0.11013$ and $0.10476$.
Since the results are ensemble averages over configurations,
the displayed error includes both the quantum fluctuations
and the error due to a finite number of noise vectors.

In order to estimate the contribution of disconnected diagrams we have
to control both errors. 
A reasonable approach is to choose the number of noise vectors,
$N_{NV}$, so as to keep both errors at similar levels.
Although we can find an ``optimal $N_{NV}$'' at each $t$ by observing
$N_{NV}$ dependence of the error displayed in Fig.~\ref{fig:Z2noise},
it is sufficient to roughly estimate such $N_{NV}$, 
since increasing the number of configurations,
$N_{conf}$, also reduces the
error of noise method and we choose rather large values of $N_{conf}$.
The number of noise vectors, $N_{NV}$, and gauge configurations,
$N_{conf}$ selected for each valence quark
mass are displayed in Table~\ref{tab:qparams}. 
As the quark mass increases, the signal to noise ratio becomes worse
rapidly. Thus a larger number of configurations as well
as of noise vectors are prepared in this calculation.

\subsection{Quarkonium correlators}
 \label{subsec:Correlators}

\FIGURE[t]{
\centerline{
\epsfig{file=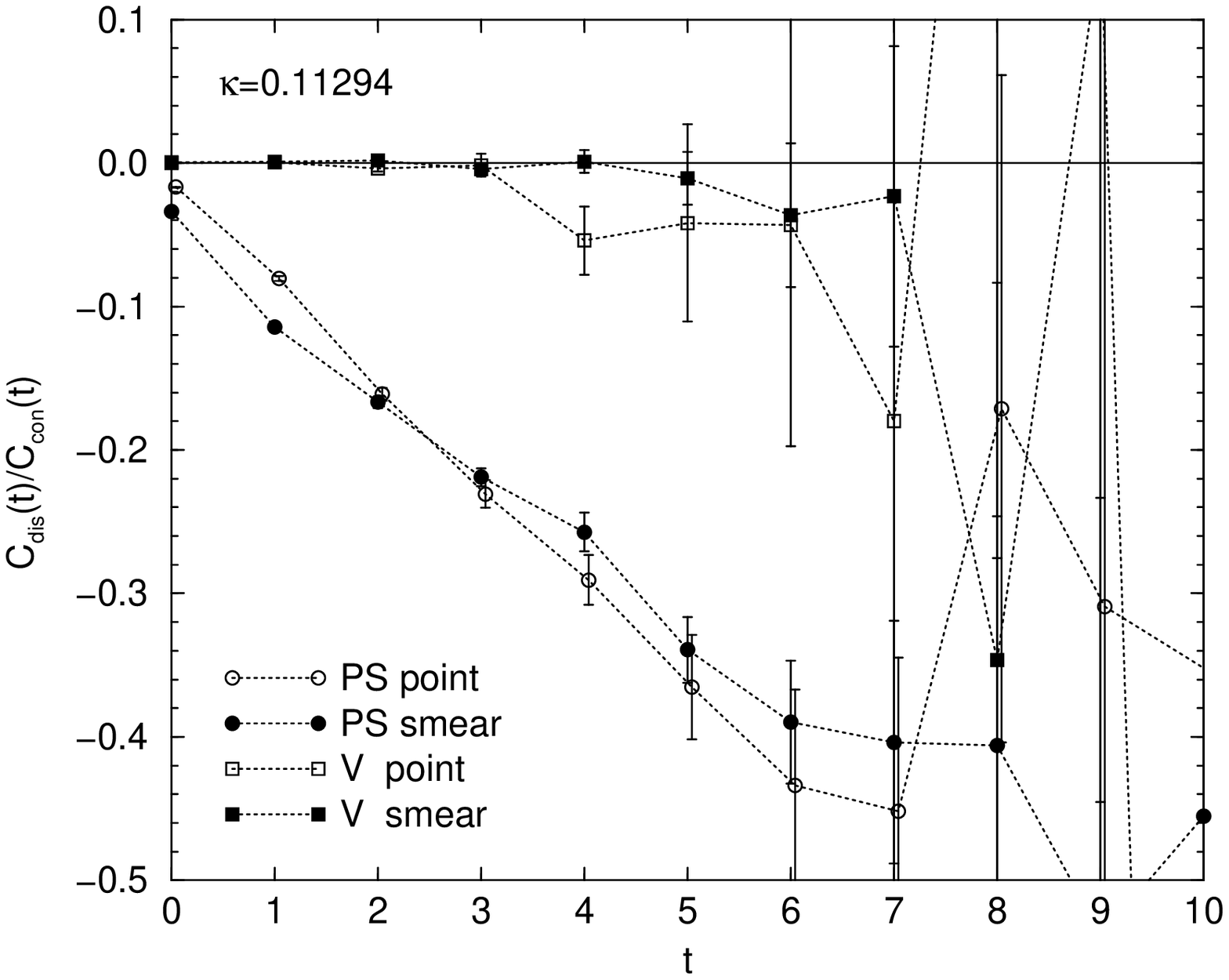,width=7cm} 
\epsfig{file=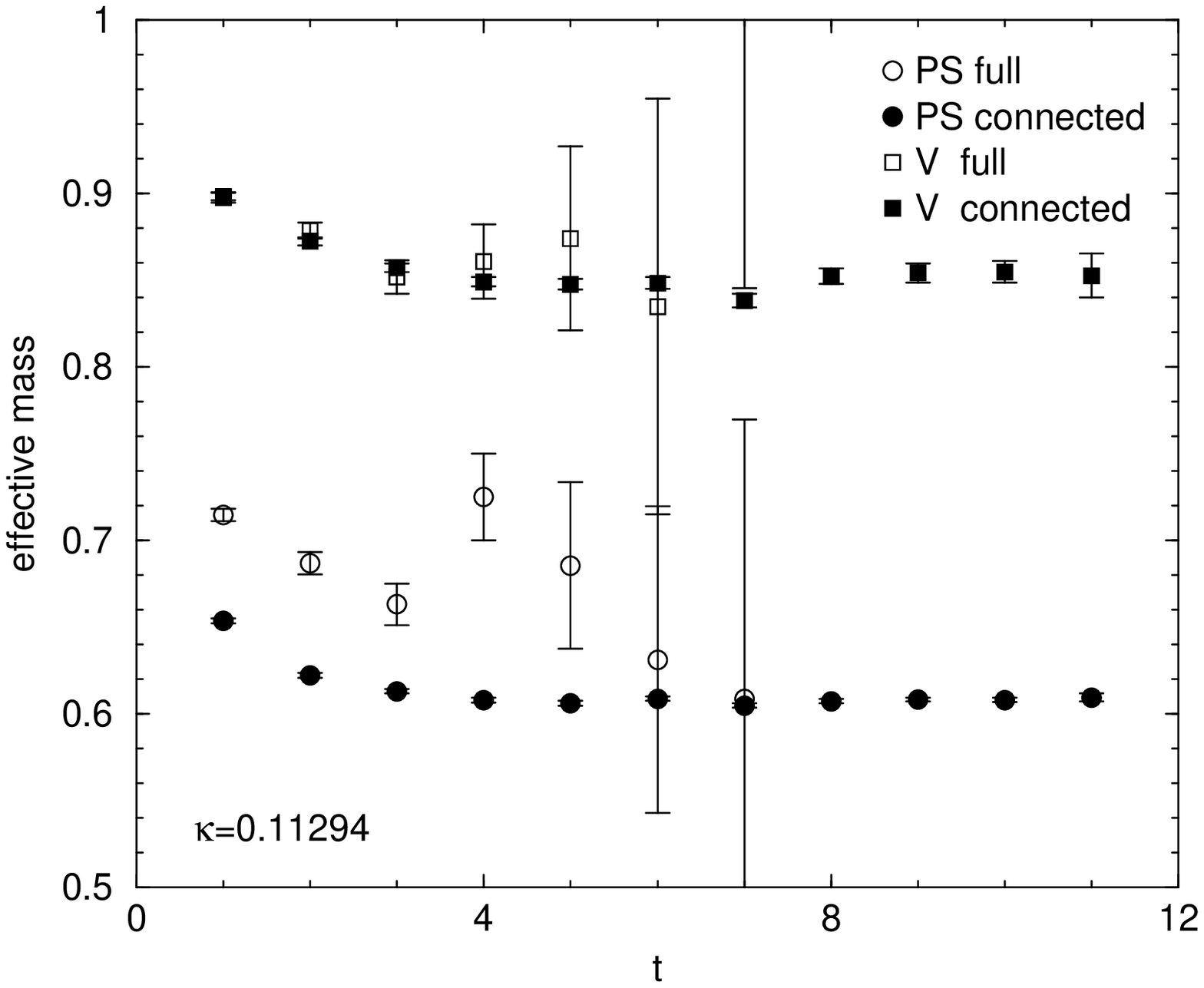,width=7cm} }
\centerline{
\epsfig{file=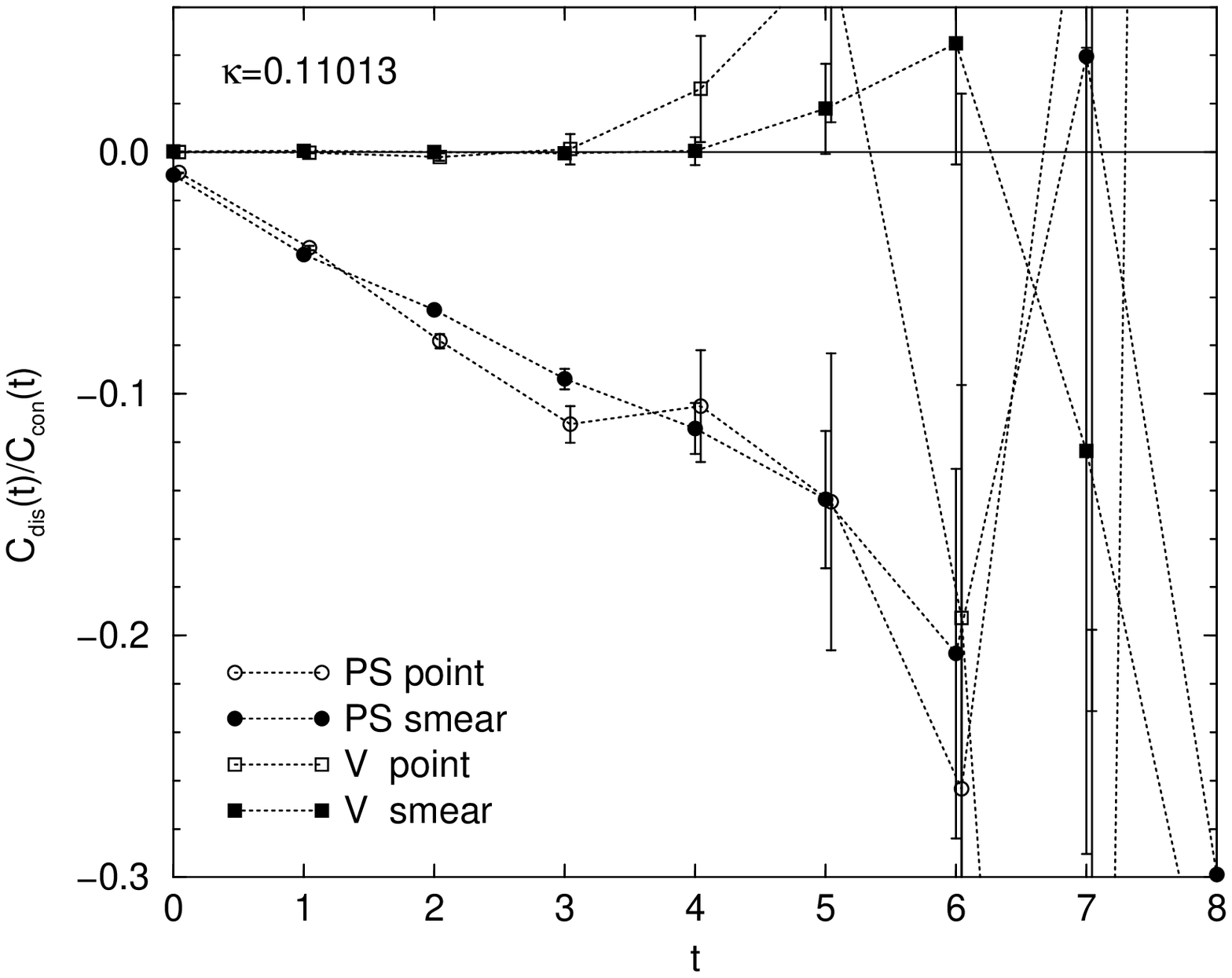,width=7cm} 
\epsfig{file=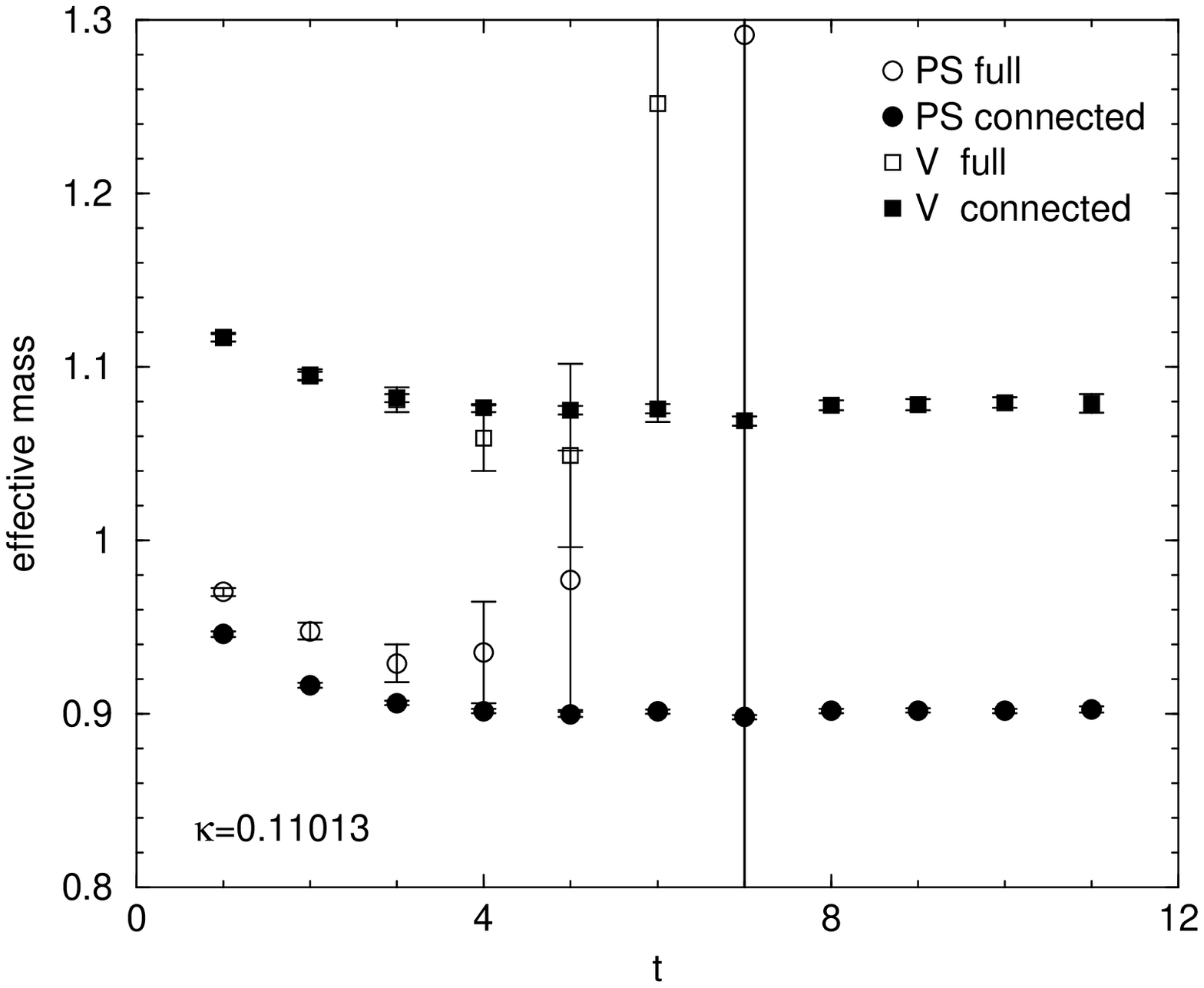,width=7cm} }
\centerline{
\epsfig{file=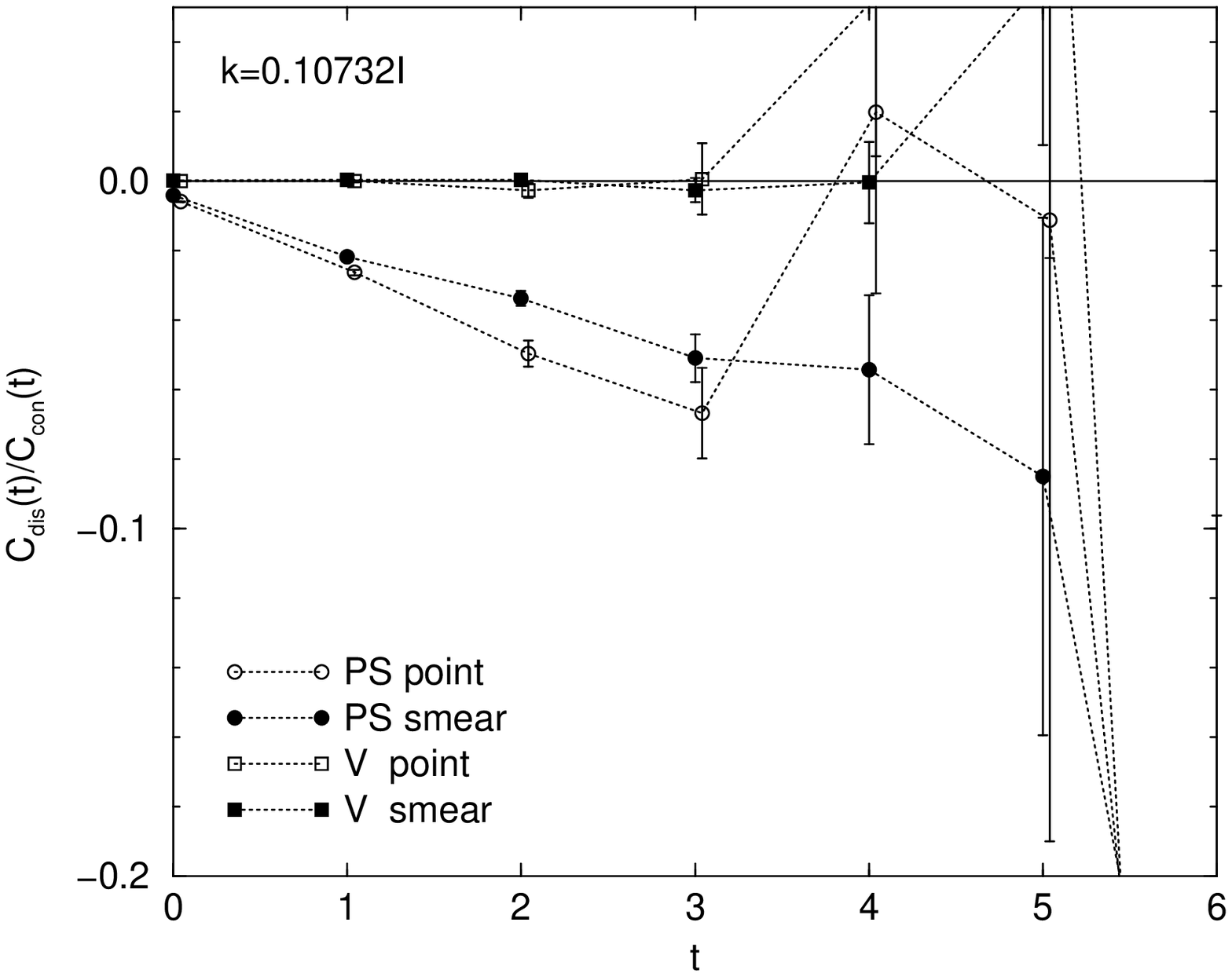,width=7cm} 
\epsfig{file=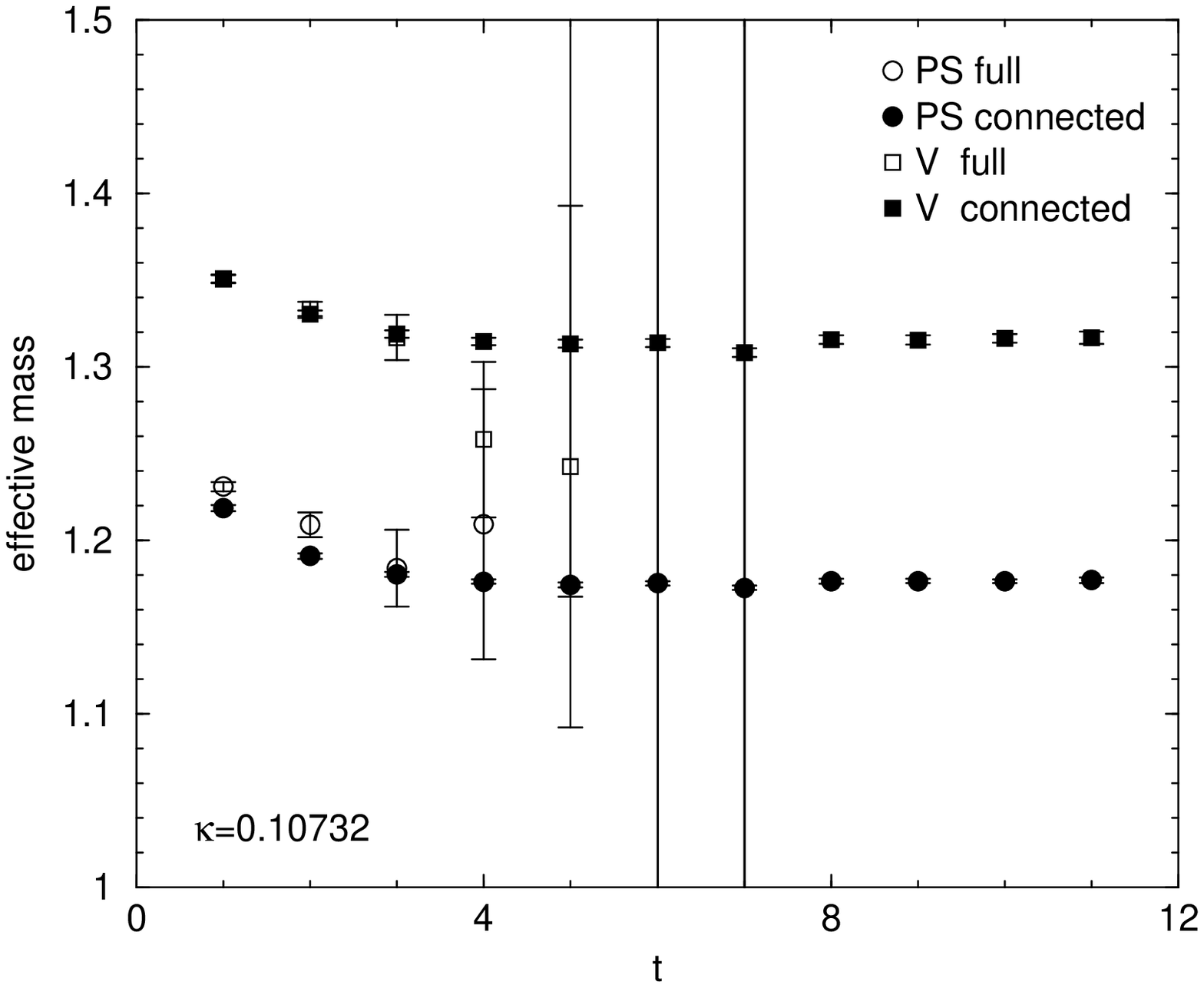,width=7cm} }
\vspace{-0.8cm}
\caption{
Ratio of disconnected to connected correlators (left panels)
and effective mass of full and connected correlators (right panels)
for $\kappa=0.11294$ (top), 0.11013 (middle), and 0.10732 (bottom).}
\label{fig:light1}
}

\FIGURE[t]{
\centerline{
\epsfig{file=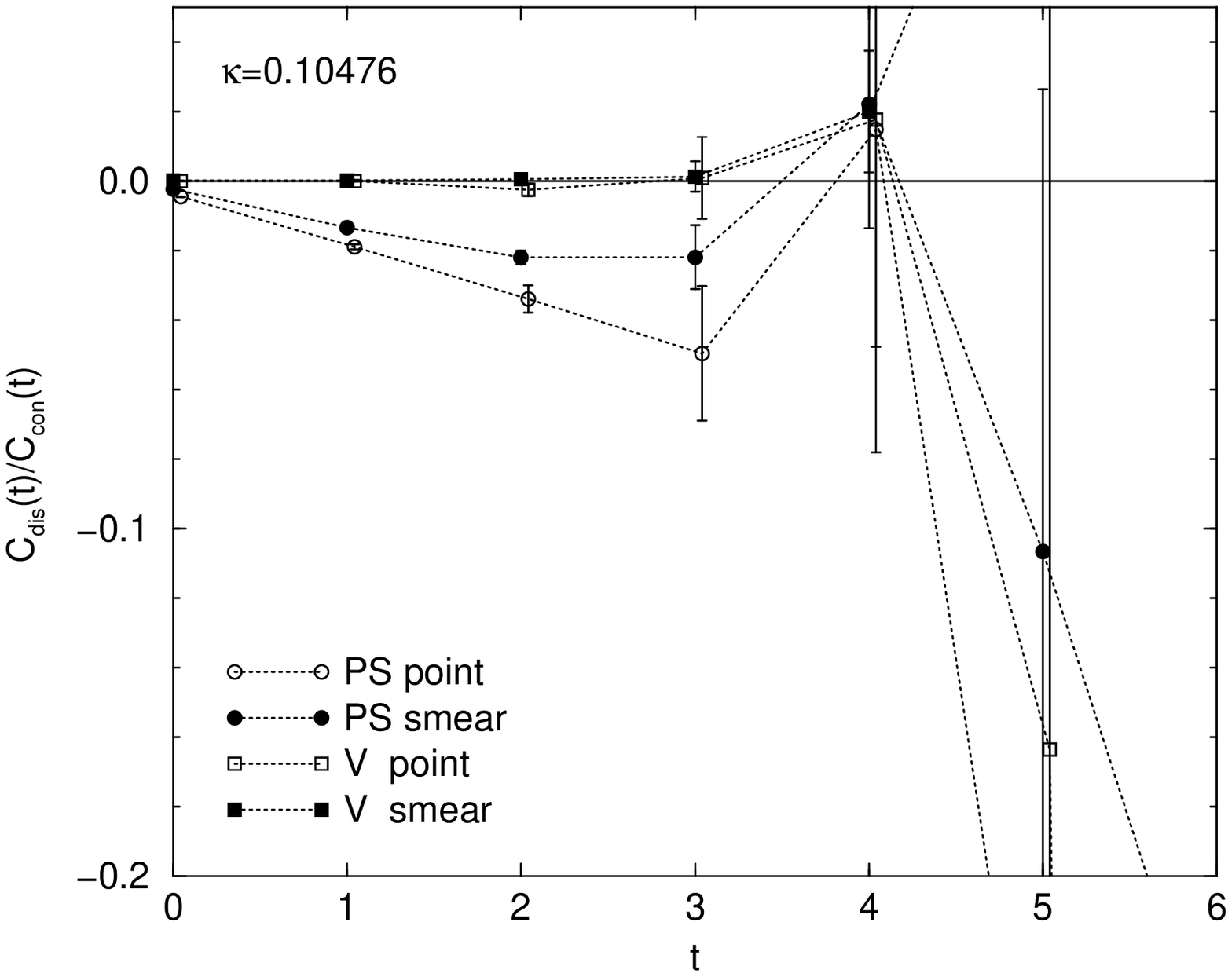,width=7cm} 
\epsfig{file=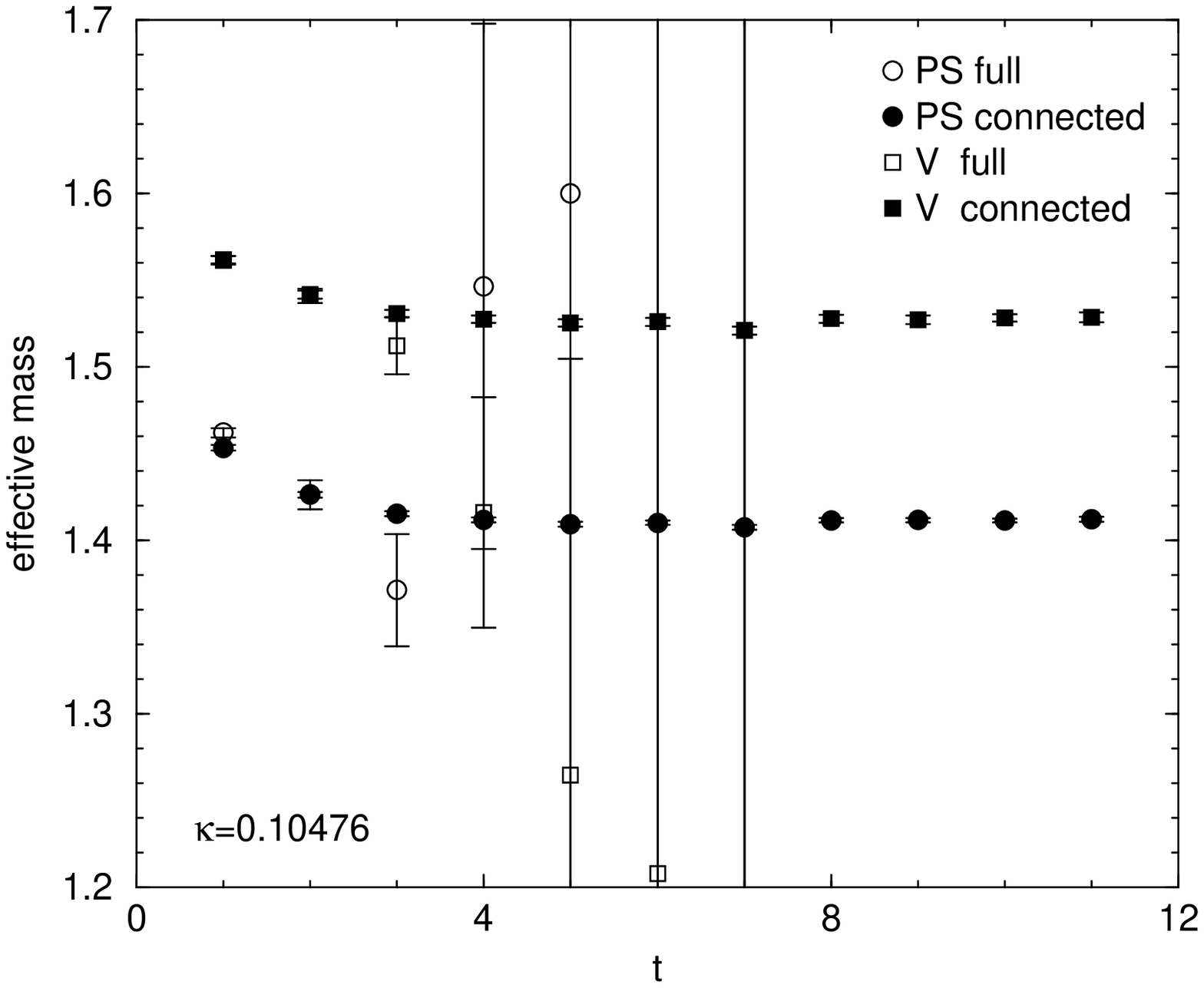,width=7cm} }
\centerline{
\epsfig{file=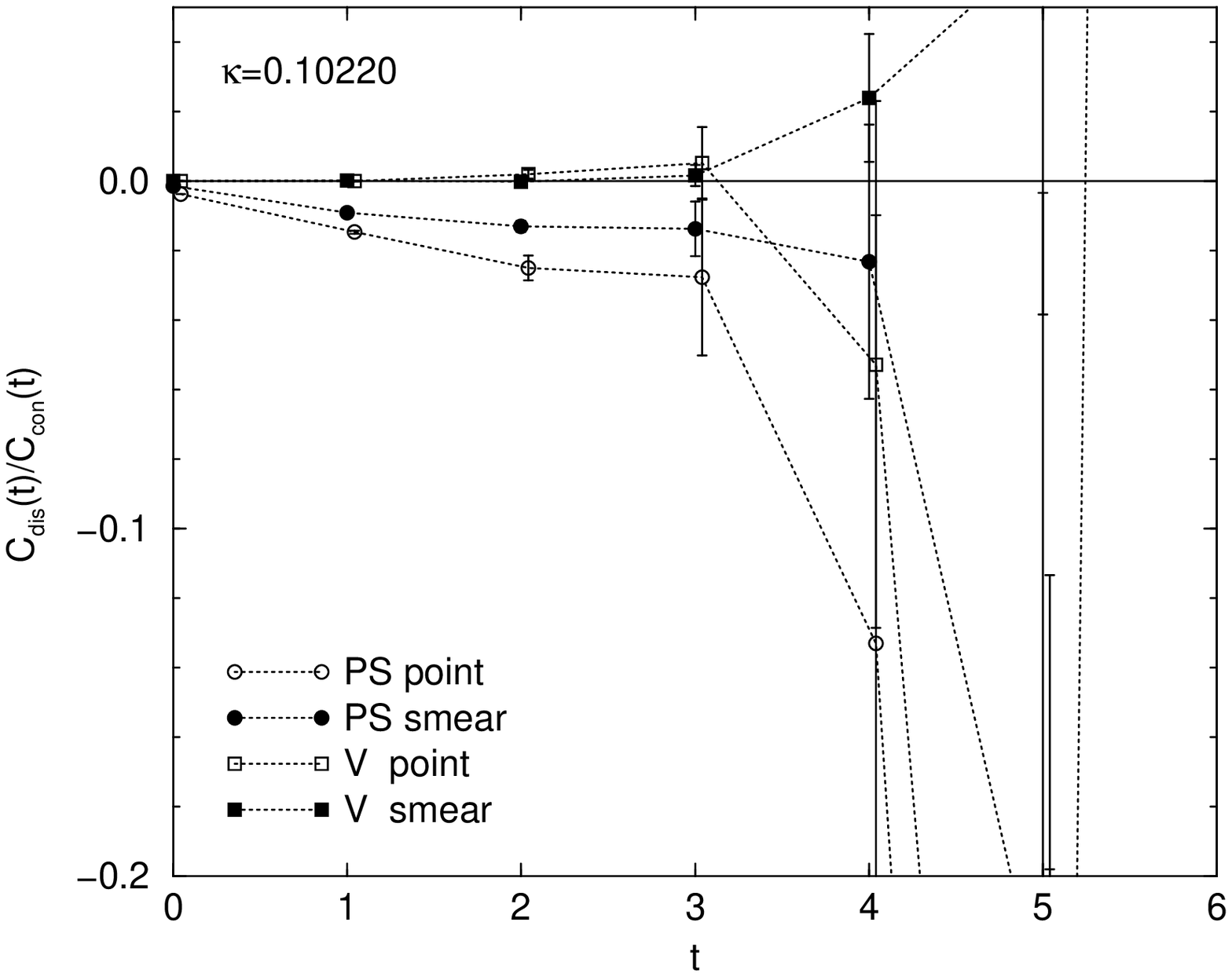,width=7cm} 
\epsfig{file=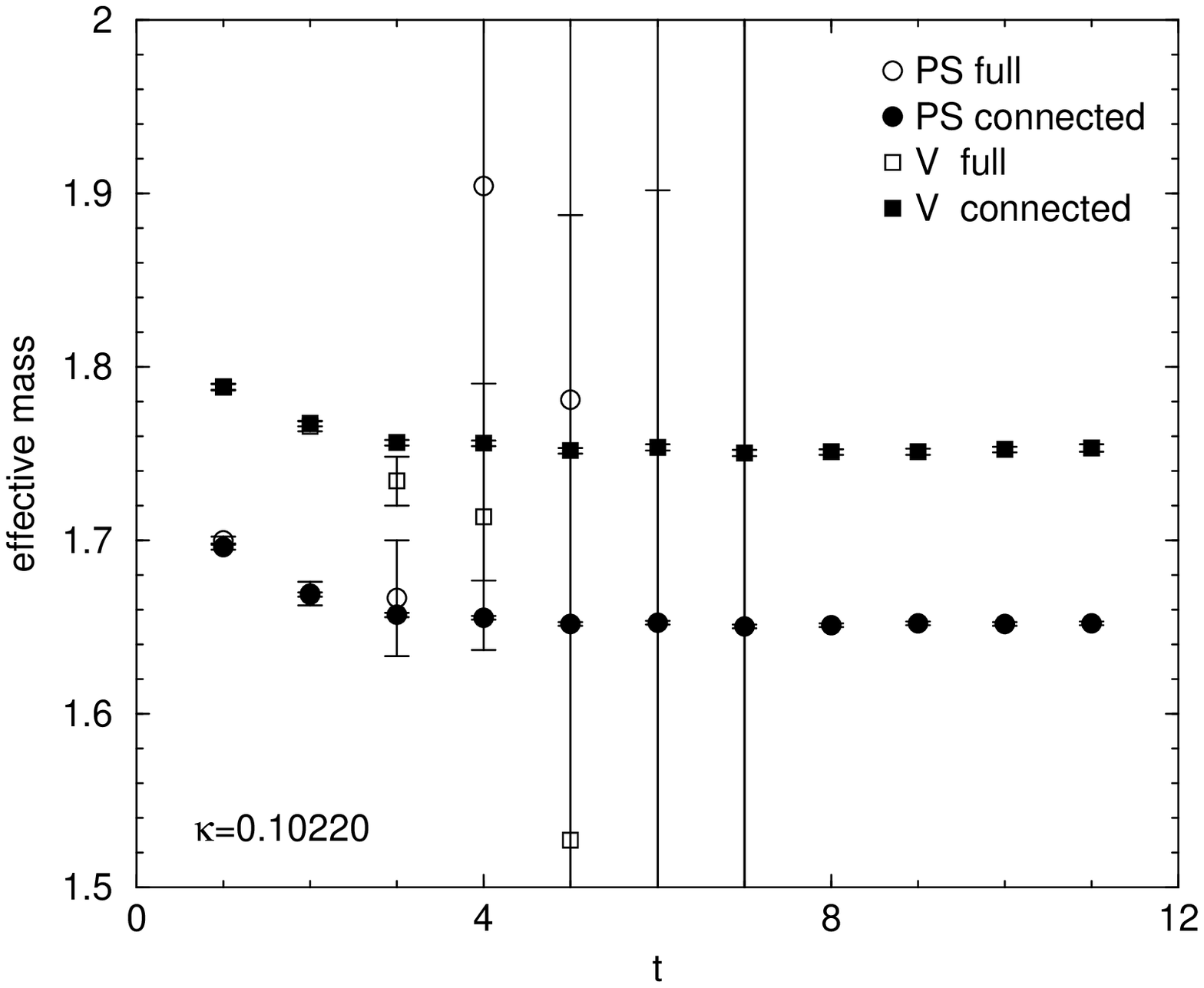,width=7cm} }
\centerline{
\epsfig{file=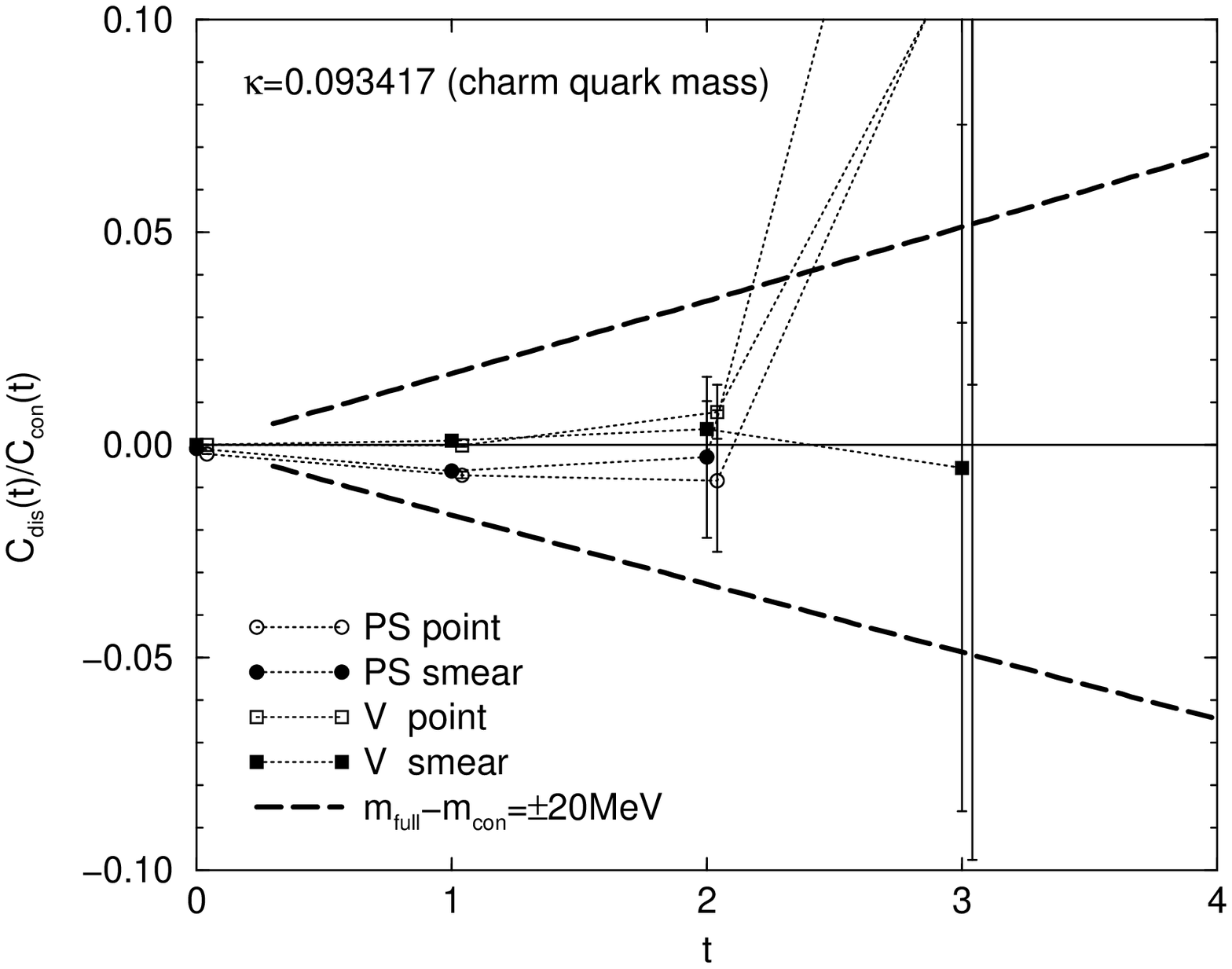,width=7cm} 
\epsfig{file=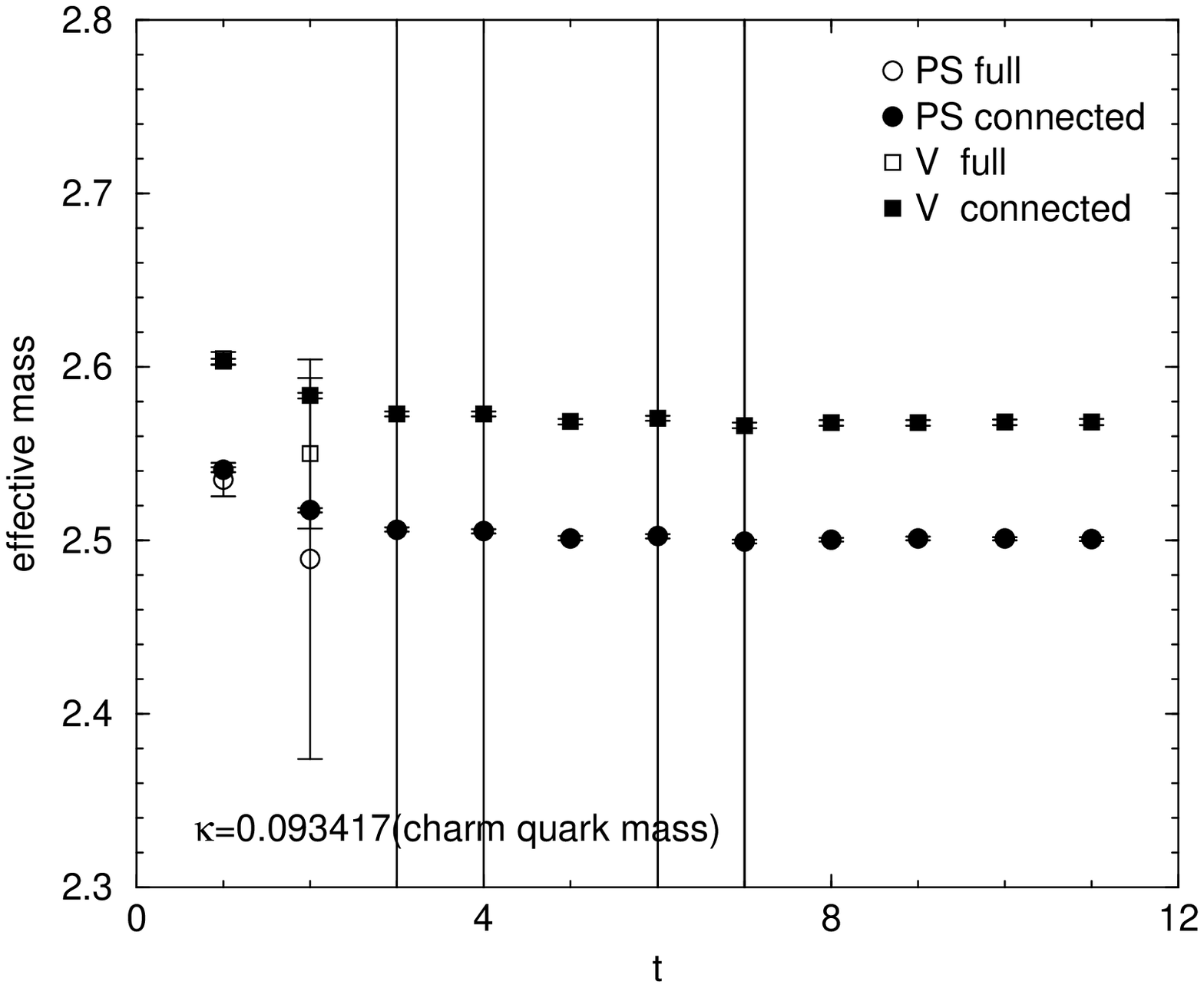,width=7cm}} 
\vspace{-0.8cm}
\caption{
The same quantities as Fig.~\ref{fig:light1} 
for $\kappa=0.10476$ (top), 0.10220 (middle) and 0.093417(bottom).
The latter value of $\kappa$ corresponds to the charm quark.
In the left bottom panel, the dashed curves
represents the cases where $m_{full}-m_{con}=\pm 20$ MeV with 
$A_{full}/A_{con}=1$.
}
\label{fig:light2}
}

In this section we evaluate the relevance of the disconnected part of 
the meson correlator for the meson spectrum.
Figs.~\ref{fig:light1} and \ref{fig:light2} show
the ratio $R(t)=C_{dis}(t)/C_{con}(t)$.
$R(t)$ is often used to evaluate the contribution of
the disconnected diagram \cite{McNeile01}.
If the ground state dominates both the connected 
and full correlators, their asymptotic time dependence can  be described
by single exponentials:
\begin{eqnarray}
C_{con}(t) &=& A_{con} \exp{(-m_{con} t)}, \\
C_{full}(t) = C_{con}(t)+C_{dis}(t)
           &=& A_{full} \exp{(-m_{full} t)}.
\end{eqnarray}
Then the ratio behaves as
\begin{equation}
R(t)= \frac{C_{dis}(t)}{C_{con}(t)}=\frac{A_{full}}{A_{con}} 
\exp{\{-(m_{full}-m_{con})t\}}-1.
\label{eq:ratio}
\end{equation} 
This ratio is useful for exploring the sign and magnitude of
the mass difference ($m_{full}-m_{con}$) when the signal is noisy.
In Figs.~\ref{fig:light1} and \ref{fig:light2}, we plot
both point and smeared correlator ratios $R(t)$.
For the latter, effective masses for full and connected
correlators are also displayed in the right panels of Figs.~\ref{fig:light1}
and \ref{fig:light2}.

For the vector channel, we find no sizable contribution from the
disconnected diagram in the whole quark mass region explored.
Similar results have been also reported in \cite{McNeile01}.
This is consistent with the OZI suppression.
In contrast, in the pseudoscalar channel, a clear signal is observed
in the light quark mass region giving rise to an increase of the pseudoscalar
mass. This is consistent with previous works 
on the flavour singlet $\eta$ ($\eta'$) meson and with theoretical 
expectations. It is important to note that this effect is opposite to
what would be needed to match the experimental result for charmonium. 
Given that the vector mass receives no sizable contribution, a decrease of 
the $\eta_c$ mass by about 30--40 MeV is required to bring the lattice 
measurement of the hyperfine splitting up to the value of 117 MeV measured 
in experiment.
Indeed, the mass difference between full and connected correlators
rapidly decreases as the quark mass increases, leaving in principle room for
a change in the sign of the mass difference.
In the left bottom panel of Fig. \ref{fig:light2}, the dashed curves represent
$R(t)$ from Eq.~\ref{eq:ratio} for $m_{full}-m_{con}=\pm 20$ MeV, assuming
$A_{full}/A_{con}=1$. In the scale
of the plot, a small deviation of $A_{full}/A_{con}$ from one would induce
a shift of the curves, keeping the slopes almost constant.
Due to the large statistical errors for $t>2$ we cannot make any definite 
statement about the slope of the data: it is compatible with zero
within errors but small slopes of order $\pm 20$ MeV cannot be ruled out.

\FIGURE[t]{
\centerline{\epsfig{file=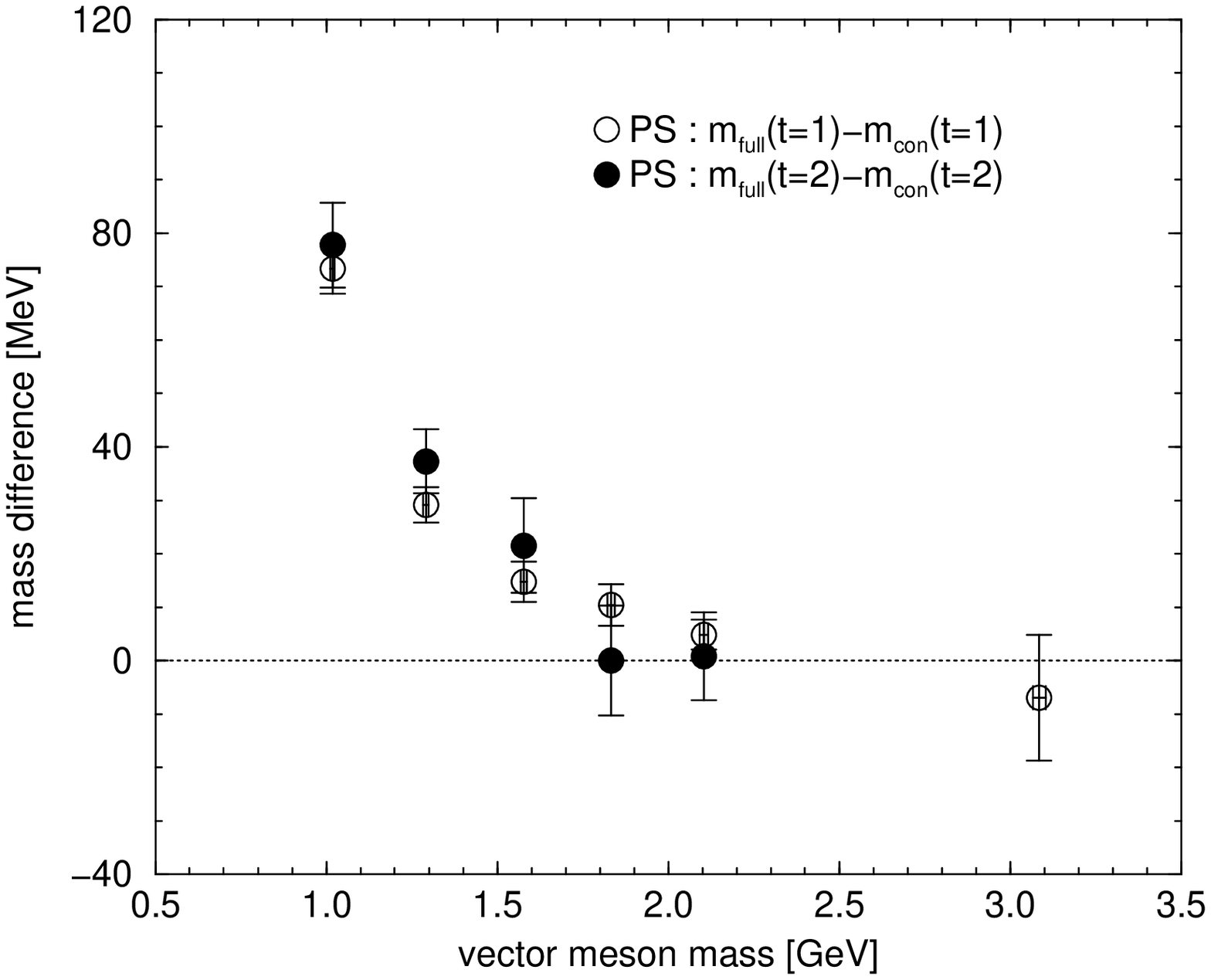,width=9cm}} 
\vspace{-0.8cm}
\caption{
The mass difference between the full and 
connected correlators, in the pseudoscalar channel, defined with the effective 
masses at $t=1$ and $t=2$ as a function of vector meson mass.
We call the effective mass at "$t=1$" that derived from comparing
$t=2$ and $t=1$.}
\label{fig:qmassdep}
}

To better settle this point, we plot in Fig.~\ref{fig:qmassdep} the mass 
difference between the full and connected correlators for the pseudoscalar 
channel as a function of the vector meson mass.
Here the mass difference is defined by the difference between
effective masses of full and connected correlators at $t=1$
and $t=2$. At these time slices, the effective mass plots in
Figs.~\ref{fig:light1} and \ref{fig:light2} do not exhibit
clear plateau behaviour. However, the contribution of the excited states 
partially cancel in the difference.
In fact, the results for $t=1$ and 2 differ from each other only
slightly. We do not plot the result for $t=2$ at the charm quark mass because
the result is too noisy. Figure~\ref{fig:qmassdep} again shows that the 
mass difference between full and connected correlator is positive for
light quark masses. It rapidly decreases as the quark mass increases,
becoming almost zero at about half the charm quark mass.
Our results at the charm quark mass are not conclusive. At the one sigma level 
they are compatible both with zero mass difference or with a negative mass
difference of about $-20$ MeV. This is consistent with the results
found by McNeile and Michael (UKQCD Collaboration) in \cite{McNeile04}.

\section{Conclusions and discussion}
 \label{sec:Conclusion}

We have investigated the contribution of disconnected diagrams to
the hyperfine splitting of quarkonium in a quark mass region ranging
from strange to charm quark masses.
As displayed in Figures~\ref{fig:light1}, \ref{fig:light2},
we found almost no contribution of disconnected diagrams to
the correlators in the vector channel in the whole quark mass region
explored in this work. For the pseudoscalar channel, however, there
is a sizable contribution around the strange quark mass region which
quickly decreases as the quark mass increases and almost vanishes
around half the charm quark mass. 
At the charm quark mass, the contribution of the disconnected
correlator is very small, in agreement with the expected OZI suppression. 
Given our large statistical errors
we cannot, however, rule out that the mass difference between full and
connected correlators becomes negative, although small.
In this respect we agree with the results previously found by 
McNeile and Michael \cite{McNeile04} who found room for a mass difference
of the order of $-20$ MeV .

To determine how large is the contribution of disconnected diagrams
for more physical situations, one needs to perform the chiral
extrapolation in the sea quarks and the continuum limit.
As argued in the introduction, going to lighter sea quarks may considerably
modify the contribution from disconnected diagrams. In particular,
an effect that could induce a further decrease in the $\eta_c$ mass would
be mixing with a pseudoscalar glueball were this to be lighter than
the $\eta_c$. This is a possibility that has been discussed also 
in \cite{McNeile04}, although it has been considered not very likely
given the fact that lattice determinations of the pseudoscalar glueball
gives masses below that of the $\eta_c$ meson.

\subsection*{Acknowledgements}

The numerical simulations were done using
a Hitachi SR8000 at the High Energy Accelerator Research
Organization (KEK),
and a NEC SX-5 at the Research Center for Nuclear Physics,
Osaka University.
It should be noted that the success of simulation owes also
to a gigabit network, SuperSINET supported by National
Institute of Informatics, for efficient and timely data transfer.
H. M. and T. U. were supported by the Japan Society for the Promotion
of Science for Young Scientists.
M.G.P. has been supported by a Ram\'on y Cajal contract of the MCyT.

\end{document}